\documentclass[aps,preprint,showkeys]{revtex4}

\usepackage{natbib}
\usepackage[dvips]{graphicx}

\usepackage{amssymb,amsfonts,amsmath}

\newcommand{\bbb}{\sffamily \bfseries}
\newcommand{\etal}{{\it et al.~}}

\begin{document}

\title[The global cargo ship network]
      {The complex network of global cargo ship movements}

\author{Pablo Kaluza}
\affiliation {Institute for Chemistry and Biology of the
  Marine Environment, Carl von Ossietzky Universit\"at, 
  Carl-von-Ossietzky-Str.~9-11, 26111 Oldenburg, Germany}

\author{Andrea K\"olzsch}
\affiliation {Institute for Chemistry and Biology of the
  Marine Environment, Carl von Ossietzky Universit\"at,
  Carl-von-Ossietzky-Str.~9-11, 26111 Oldenburg, Germany}

\author{Michael T. Gastner}
\affiliation {Institute for Chemistry and Biology of the
  Marine Environment, Carl von Ossietzky Universit\"at,
  Carl-von-Ossietzky-Str.~9-11, 26111 Oldenburg, Germany}

\author{Bernd Blasius}
\email{blasius@icbm.de}
\affiliation {Institute for Chemistry and Biology of the
  Marine Environment, Carl von Ossietzky Universit\"at,
  Carl-von-Ossietzky-Str.~9-11, 26111 Oldenburg, Germany}

\begin{abstract} 
  Transportation networks play a crucial role in human mobility, the 
  exchange of goods, and the spread of invasive species.  
  With 90\% of world trade carried by sea, the global network of merchant 
  ships provides one of the most important modes of transportation.  
  Here we use information about the itineraries of 16,363 cargo ships during 
  the year 2007 to construct a network of links between ports.  
  We show that the network has several features which set it apart from 
  other transportation networks.  
  In particular, most ships can be classified in three categories: bulk dry 
  carriers, container ships and oil tankers.  
  These three categories do not only differ in the ships' physical 
  characteristics, but also in their mobility patterns and networks.  
  Container ships follow regularly repeating paths whereas bulk dry 
  carriers and oil tankers move 
  less predictably
  between ports.  
  The network of all ship movements possesses a heavy-tailed distribution 
  for the connectivity of ports and for the loads transported on the links 
  with systematic differences between ship types.  
  The data analyzed in this paper improve current assumptions based on
  gravity models of ship movements, an important step towards understanding
  patterns of global trade and bioinvasion.
\end{abstract}


\maketitle

\section{Introduction}

The ability to travel, trade commodities, and share information 
around the world with unprecedented efficiency is a defining feature of the 
modern globalized economy.  Among the different means of transport, ocean 
shipping stands out as the most energy efficient mode of long-distance 
transport for large quantities of goods (Rodrigue \etal 2006).  According 
to estimates, as much as 90\% of world trade is hauled by ships (International
Maritime Organization 2006).  
In 2006, 7.4 billion tons of goods were loaded at the world's ports.  
The trade volume currently exceeds 30 trillion ton-miles 
and is growing at a rate faster than the global economy (United Nations
conference on trade and development 2007). 

The worldwide maritime network also plays a crucial role in today's spread 
of invasive species.  Two major pathways for marine bioinvasion are 
discharged water from ships' ballast tanks (Ruiz \etal 2000) and hull 
fouling (Drake \& Lodge 2007).
Even terrestrial species such as insects are sometimes inadvertently 
transported in shipping containers (Lounibos 2002). 
In several parts of the world, invasive species have caused dramatic levels 
of species extinction and landscape alteration, thus damaging ecosystems and
creating hazards for human livelihoods, health, and 
local economies (Mack \etal 2000). 
The financial loss due to bioinvasion is estimated to be \$120 
billion per year in the United States alone (Pimentel \etal 2005).

Despite affecting everybody's daily lives, the shipping industry is far 
less in the public eye than other sectors of the global transport 
infrastructure. 
Accordingly, it has also received little attention in the recent literature 
on complex networks (Wei \etal 2007, Hu \& Zhu 2009). This neglect is 
surprising 
considering the current interest in networks (Albert \& Barabasi 2002,
Newman 2003a, Gross \& Blasius 2008), 
especially airport (Barrat \etal 2004, Guimer\`a \& Amaral 2004,
Hufnagel \etal 2004, Guimer\`a \etal 2005), 
road (Buhl \etal 2006, Barthelemy \& Flammini 2008) and 
train networks (Latora \& Marchiori 2002, Sen \etal 2003).  
In the spirit of current network research, we take here a 
large-scale
perspective on the global cargo ship network (GCSN) as a complex system 
defined as the network of ports that are connected by links if ship traffic 
passes between them.  

Similar research in the past had to make strong assumptions about flows on 
hypothetical networks with connections between all pairs of ports in order 
to approximate ship movements (Drake \& Lodge 2004, Tatem \etal 2006).  
By contrast, our analysis is based on comprehensive data of real ship 
journeys allowing us to construct the actual network.
We show that it has a small-world topology where the combined
cargo capacity of ships calling at a given port 
(measured in gross tonnage)
follows a heavy-tailed distribution. 
This capacity scales superlinearly with the number of directly connected 
ports.  We identify the most central ports in the network and find several 
groups of highly interconnected ports showing the importance of regional 
geopolitical and trading blocks.

A high-level description of the complete network, however, does not yet
fully capture the network's complexity. Unlike previously studied 
transportation networks, the GCSN has a multi-layered structure.   
There are, broadly speaking, three classes of cargo ships -- container 
ships, bulk dry carriers, and oil tankers --  that span distinct subnetworks.
Ships in different categories tend to call at different ports and travel in 
distinct patterns.  
We analyze the trajectories of individual ships in the GCSN and develop 
techniques to extract quantitative information about characteristic movement 
types.  
With these methods we can quantify that container ships sail along more 
predictable, frequently repeating routes than oil tankers or bulk dry 
carriers.  
We compare the empirical data with theoretical traffic flows calculated by the 
gravity model.   
Simulation results, based on the full GCSN data or the gravity model
differ significantly in a
population-dynamic
model for the spread of invasive species between the world's ports.  
Predictions based on the real network are thus more informative for 
international policy decisions concerning the stability of worldwide trade 
and for reducing the risks of bioinvasion.

\section{Data}

An analysis of global ship movements requires detailed knowledge of 
ships' arrival and departure times at their ports of call.  Such data
have become available in recent years.  Starting in 2001, ships and ports 
have begun installing Automatic Identification System (AIS) equipment.  
AIS transmitters on board of the ships automatically report the arrival and 
departure times to the port authorities.  This technology is primarily used 
to avoid collisions and increase port security, but arrival and departure 
records are also made available by Lloyd's Register Fairplay for commercial 
purposes as part of its Sea-web data base (www.sea-web.com).  
AIS devices have not been installed in all ships and ports yet, and 
therefore there are some gaps in the data.  
Still, all major ports and the largest ships are included, thus the data 
base represents the majority of cargo transported on ships.

\begin{figure*}[t]
  \begin{center}
    \includegraphics[width=\textwidth,clip]{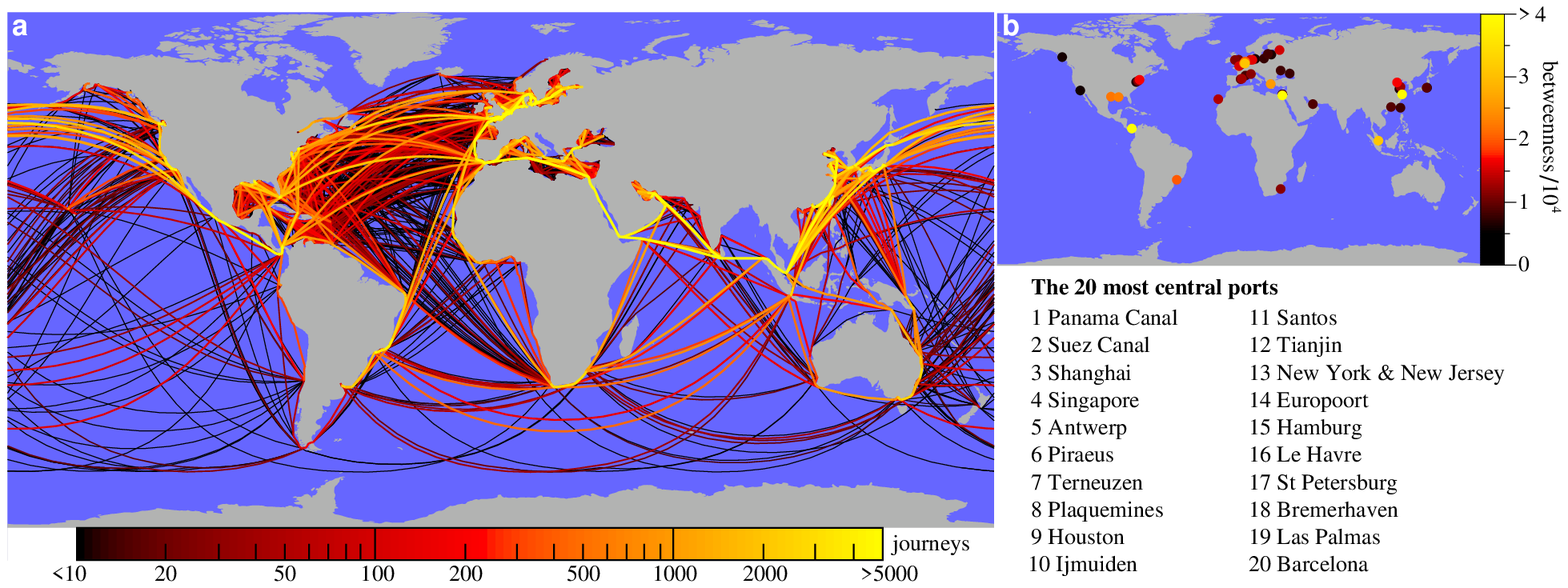}
    \caption{\label{full_netw}
      Routes, ports and betweenness centralities in the global cargo ship 
      network (GCSN).
      (a) The trajectories of all cargo ships bigger than $10,000$ GT 
      during 2007.  The color scale indicates the number of journeys along 
      each route.
      Ships are assumed to travel along the shortest (geodesic) paths on water.
      (b) A map of the 50 ports of highest betweenness centrality
      and a ranked list of the 20 most central ports.
    }
  \end{center}
\end{figure*}

Our study is based on Sea-web's arrival and departure records in the calendar 
year 2007 as well as Sea-Web's comprehensive data on the ships' physical 
characteristics. 
We restrict our study to cargo ships bigger than $10,000$ GT (gross tonnage) 
which make up 93\% of the world's total capacity for ship cargo transport.
From these we select all $16,363$ ships for which AIS data are available,
taken as representative of the global traffic and long-distance trade 
between the $951$ ports equipped with AIS receivers (for details see 
Electronic Supplementary Material).  
For each ship we obtain a trajectory from the data base, i.e. a list of 
ports visited by the ship sorted by date. 
In 2007, there were $490,517$ nonstop journeys linking $36,351$ distinct 
pairs of arrival and departure ports.  The complete set of trajectories, each 
path representing the shortest route at sea and colored by the number of 
journeys passing through it, is shown in Fig.~\ref{full_netw}a.

Each trajectory can be interpreted as a small directed network where the 
nodes are ports linked together if the ship traveled directly between the 
ports.
Larger networks can be defined by merging trajectories of different ships. 
In this article we aggregate trajectories in four different ways:
the combined network of all available trajectories, and the subnetworks of 
container ships ($3\,100$ ships), bulk dry carriers ($5\,498$) 
and oil tankers ($2\,628$).
These three subnetworks combined cover 74\% of the GCSN's total gross tonnage.
In all four networks, we assign a weight $w_{ij}$ to the link from port $i$ 
to $j$ equal to the sum of the available space on all ships that have 
traveled on the link during 2007 measured in GT.  If a ship made the 
journey from $i$ to $j$ more than once, its capacity contributes multiple 
times to $w_{ij}$.

\section{The global cargo ship network}

The directed network of the entire cargo fleet is noticeably asymmetric, with
59\%  of all linked pairs of ports being connected only in one direction. 
Still, the vast majority of ports (935 out of 951) belongs to one single 
strongly connected component, i.e. for any two ports in this component 
there are routes in both directions, though possibly visiting different 
intermediate ports.
The routes are intriguingly short: only few steps in the network are needed 
to get from one port to another.  The shortest path length $l$ between two 
ports is the minimum number of nonstop connections one must take to travel 
between origin and destination.
In the GCSN, the average over all pairs of ports is extremely small, 
$\langle l\rangle = 2.5$.  Even the maximum shortest path between any two 
ports (e.g. from Skagway, Alaska, to the small Italian island of Lampedusa),
is only of length $l_\text{max} = 8$.  In fact, the majority of all possible 
origin-destination pairs (52\%) can already be connected by two steps or less.

Comparing these findings to those reported for the worldwide airport 
network (WAN) shows interesting differences and similarities.
The high 
asymmetry 
of the GCSN has not been found in the WAN, indicating that 
ship traffic is structurally very different from 
aviation.
Rather than being formed by the accumulation of back and forth trips, ship 
traffic seems to be governed by an optimal arrangement of unidirectional, 
often circular routes.  This optimality also shows in the GCSN's small 
shortest path lengths.  In comparison,
in the WAN, 
the average and maximum shortest path lengths are $\langle l\rangle = 4.4 $ 
and $l_\text{max} = 15$ respectively (Guimer\`a \etal 2005), i.e. about twice
as long as in the GCSN.  
Similar to the WAN, the GCSN is highly clustered: if a port $X$ is linked to 
ports $Y$ and $Z$, there is a high probability that there is also a 
connection from $Y$ to $Z$.  
We calculated a clustering coefficient $C$ (Watts \& Strogatz 1998) for 
directed networks and find $C=0.49$ whereas random networks with the same 
number of nodes and links only yield $C=0.04$ on average.  
%AK 
Degree dependent clustering coefficients $C_k$ reveal that clustering decreases with node degree (see Electronic Supplementary Material).
%\AK
Therefore, the GCSN --~like the WAN~-- can be regarded as a small-world 
network possessing short path lengths despite substantial 
clustering (Watts \& Strogatz 1998).  
However, the average degree of the GCSN, i.e. the average number of links 
arriving at and departing from a given
port (in- plus out-degree),
$\langle k\rangle= 76.5$, is notably higher than in the WAN where 
$\langle k \rangle = 19.4 $ (Barrat \etal 2004).
In the light of the network size (the WAN consists of 3880 nodes), this 
difference becomes even more pronounced, indicating that the GCSN is much more 
densely connected.  This redundancy of links gives the network high structural 
robustness to the loss of routes for keeping up trade.

\begin{figure*}[tb] 
  \begin{center}
    \includegraphics[width=0.7\textwidth,clip]{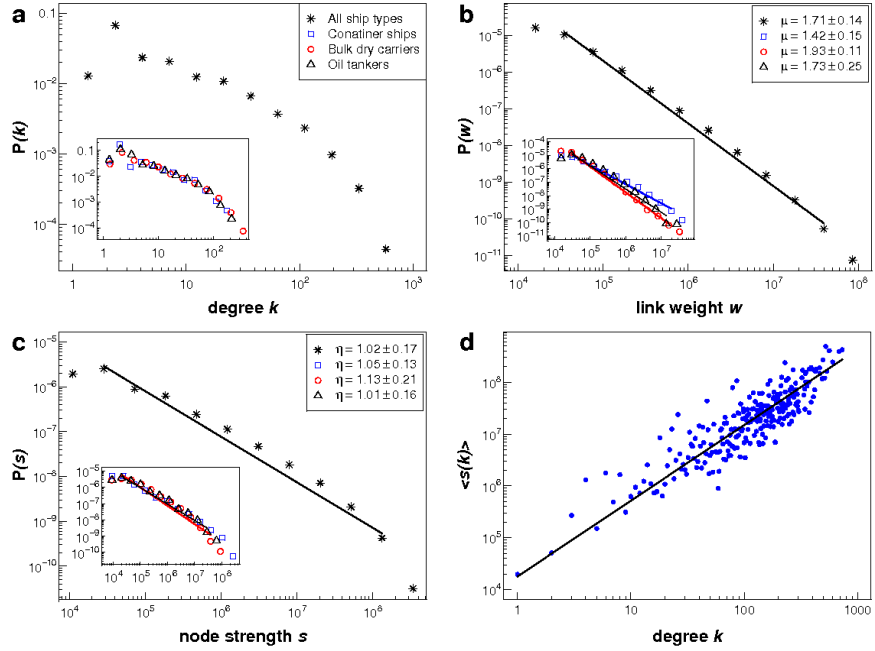}
    \caption{\label{degree_distribution} 
      Degrees and weights in the global cargo ship network $*$ (insets: 
      subnetworks for container ships {\tiny $\Box$}, bulk dry carriers 
      $\circ$, and oil tankers {\tiny $\triangle$}).
      (a) The degree distributions $P(k)$ are right-skewed, but not 
      power laws, neither for the GCSN nor its subnetworks.  The degree $k$ 
      is defined here as the sum of in- and out-degree, thus $k=1$ is rather 
      rare.
      (b) The link weight distributions $P(w)$ reveal clear power 
      law relationships for the GCSN and the three subnetworks, with 
      exponents $\mu$ characteristic for the movement patterns of the 
      different ship types. 
      (c) The node strength distributions $P(s)$ are also heavy-tailed, 
      showing power law relationships. The stated exponents are calculated 
      by linear regression with $95\%$ confidence intervals (similar results 
      are obtained with maximum likelihood estimates, see Electronic 
      Supplementary Material). 
      (d) The average strength of a node $\langle s(k) \rangle$ scales 
      superlinearly with its degree,
      $\langle s(k) \rangle \propto k^{1.46\pm0.1}$, indicating that highly 
      connected ports have, on average, links of higher weight.
    }
  \end{center}
\end{figure*}

The degree distribution $P(k)$ shows that most ports have few connections, 
but there are some ports linked to hundreds of other ports 
(Fig.~\ref{degree_distribution}a).  
Similar right-skewed degree distributions have been observed in many 
real-world networks (Barabasi \& Albert 1999).
While the GCSN's degree distribution is not exactly scale-free, the
distribution of link weights, $P(w)$, follows approximately a power law 
$P(w)\propto w^{-\mu}$ with $\mu=1.71\pm 0.14$
(95\% CI for linear regression, 
Fig.~\ref{degree_distribution}b, see also Electronic Supplementary Material).  
By averaging the sums of the link weights arriving at and departing from 
port $i$, we obtain the node strength $s_i$ (Barrat \etal 2004).  
The strength distribution can also be approximated by a power law 
$P(s)\propto s^{-\eta}$ with $\eta = 1.02 \pm 0.17$,
meaning that a small number of ports handle huge amounts of cargo 
(Fig.~\ref{degree_distribution}c).
The determination of power law relationships by line fitting has been 
strongly criticised (e.g. Newman 2005, Clauset \etal 2009), therefore we 
analysed the distributions with model selection by Akaike weights (Burnham \&
Anderson 1998). Our results confirm that a power law is a better fit than an exponential or a log-normal distribution for $P(w)$ and $P(s)$, 
but not $P(k)$ (see Electronic Supplementary Material).
These findings agree well with the concept of hubs-spokes networks (Notteboom 2004) that were 
proposed for cargo traffic, for example in Asia (Robinson 1998).
There are a few large, highly connected ports through 
which all smaller ports transact their trade.  
This scale-free property makes the ship trade network prone to the spreading and persistence of bioinvasive organisms
(e.g. Pastor-Satorras \& Vespignani 2001).
The average nearest neighbors's degrees, a measure of network assortativity, additionally underline the hubs-spokes property of cargo ship traffic (see Electronic Supplementary Material).

Strengths and degrees of the ports are related according to the 
scaling relation $\langle s(k)\rangle\propto k^{1.46\pm 0.1}$ (95\% CI for 
SMA regression, Warton \etal 2006).  
Hence, the strength of a port grows generally faster than its degree 
(Fig.~\ref{degree_distribution}d). In other words, highly connected ports not 
only have many links, but their links also have a higher than average weight.  
This observation agrees with the fact that busy ports are better equipped to 
handle large ships with large amounts of cargo. 
A similar result, $\langle s(k)\rangle\propto k^{1.5\pm0.1}$, was found for 
airports (Barrat \etal 2004), which may hint at a general pattern in
transportation networks.
In the light of bioinvasion, these results underline empirical findings that 
big ports are more heavily invaded because of increased propagule pressure by 
ballast water of more and larger ships (Mack \etal 2000, Williamson 1996, see 
e.g. Cohen \& Carlton 1998).

A further indication of the importance of a node is its betweenness 
centrality (Freeman 1979, Newman 2004).  
The betweenness of a port is the number of topologically shortest directed
paths in the network that pass through this port.
In Fig.~\ref{full_netw}b we plot and list the most central ports. 
Generally speaking, centrality and degree are strongly correlated 
(Pearson's correlation coefficient: $0.81$), but in individual cases other
factors can also play a role.  The Panama and Suez Canal, for instance,
are shortcuts to avoid long passages around South America and Africa.  
Other ports have a high centrality because they are visited by a large 
number of ships (e.g. Shanghai) whereas others gain their status primarily 
by being connected to many different ports (e.g. Antwerp).

\section{The network layers of different ship types}

To compare the movements of cargo ships of different types, separate 
networks were generated for each of the three main ship types: container 
ships, bulk dry carriers, and oil tankers.  
Applying the network parameters introduced in the previous section to these 
three subnetworks reveals some broad-scale differences (see 
Table~\ref{tab:trajs}).  
The network of container ships is densely clustered, $C=0.52$, has a rather 
low mean degree, $\langle k\rangle=32.44$, and a large mean number of journeys
(i.e. number of times any ship passes) per link, $\langle J\rangle=24.26$.  
The bulk dry carrier network, on the other hand, is less clustered, has a 
higher mean degree, and fewer journeys per link ($C=0.43$, 
$\langle k\rangle=44.61$, $\langle J\rangle=4.65$).  
For the oil tankers, we find intermediate values ($C=0.44$, 
$\langle k\rangle=33.32$, $\langle J\rangle=5.07$).
Note that the mean degrees $\langle k\rangle$ of the subnetworks are 
substantially smaller than that of the full GCSN, indicating that different 
ship types use essentially the same ports but different connections.

A similar tendency appears in the scaling of the link weight distributions 
(Fig.~\ref{degree_distribution}b).    
$P(w)$ can be approximated as power laws for each network, but with different 
exponents $\mu$.
The container ships have the smallest exponent ($\mu=1.42$) and bulk dry 
carriers the largest ($\mu=1.93$) with oil tankers in between ($\mu=1.73$). 
In contrast, the exponents for the distribution of node strength $P(s)$ are 
nearly identical in all three subnetworks, $\eta=1.05$, $\eta=1.13$ and 
$\eta=1.01$, respectively.

These numbers give a first indication that different ship types move in
distinctive patterns.  
Container ships typically follow set schedules visiting several ports in a 
fixed sequence along their way, thus providing regular services.  
Bulk dry carriers, by contrast, appear 
less predictable
as they frequently change their routes on short notice depending on the 
current supply and demand of the goods they carry.  
The larger variety of origins and destinations in the bulk dry carrier 
network ($n=616$ ports, compared to $n=378$ for container ships) explains the 
higher average degree and the smaller number of journeys for a given link.  
Oil tankers also follow short-term market trends, but, because they can only 
load oil and oil products, the number of possible destinations ($n=505$) is 
more limited than for bulk dry carriers.

These differences are also underlined by the betweenness centralities of the three network layers (see Electronic Supplementary Material).  
While some ports rank 
highly in all categories (e.g. Suez Canal, Shanghai), others are specialized 
on certain ship types.  
For example, the German port of Wilhelmshaven ranks tenth in terms of its 
world-wide betweenness for oil tankers, but is only 241st for bulk dry 
carriers and 324th for container ships.

\begin{figure}[tb] 
  \begin{center}
    \includegraphics[angle=-90,width=0.9\textwidth,clip]{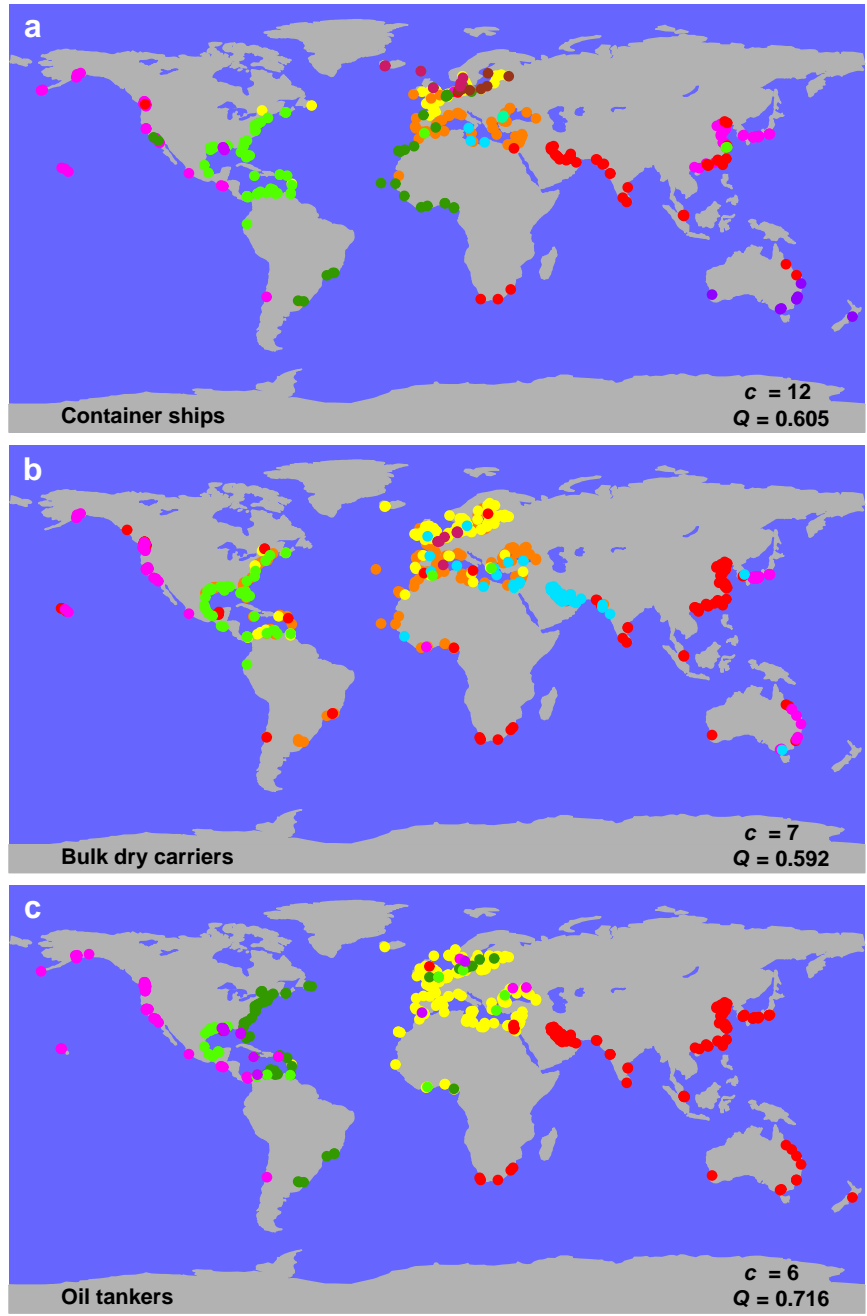}
    \caption{\label{Modularity} Communities of ports 
      in three cargo ship subnetworks.
      The communities are groups of ports that maximize the number of links 
      within the groups, as opposed to between the groups, in terms of the 
      modularity $Q$ (Leicht \& Newman 2008).  In each map, the colors 
      represent the $c$ distinct trading communities for the goods 
      transported by 
      (a) container ships, 
      (b) bulk dry carriers, and 
      (c) oil tankers.
      The optimal values for $c$ and $Q$ are stated in the lower right corners.
All modularities $Q$ of the examined networks differ significantly from modularities in Erd\"{o}s-Renyi graphs of the same size and number of links 
(Guimer\`a \etal 2004). For the networks corresponding to (a), (b) and (c) values are $Q_{ER} = 0.219$,  $Q_{ER} = 0.182$ and $Q_{ER} = 0.220$, respectively.
    }
  \end{center}
\end{figure}

We can gain further insight into the roles of the ports by examining their
community structure.  Communities are groups of ports with many links
within the groups but few links between different groups.  We calculated
these communities for the three subnetworks with a modularity optimization 
method for directed networks (Leicht \& Newman 2008)
and found that they differ significantly from modularities of corresponding 
Erd\"{o}s-Renyi graphs (Fig.~\ref{Modularity}, Guimer\'a \etal 2004).
The network of container trade shows 12 communities (Fig.~\ref{Modularity}a).
The largest ones are located
(1) on the Arabian, Asian, and South African coasts,
(2) on the North American east coast and in the Caribbean,
(3) in the Mediterranean, the Black Sea, and on the European west coast,
(4) in Northern Europe, and
(5) in the Far East and on the American west coast.
The transport of bulk dry goods reveals 7 groups (Fig.~\ref{Modularity}b). 
Some can be interpreted as geographic entities (e.g. North American east 
coast, trans-Pacific trade) while others are dispersed on multiple
continents.  Especially interesting is the community structure of the oil 
transportation network which shows 6 groups (Fig.~\ref{Modularity}c):
(1) the European, north and west African market 
(2) a large
community comprising Asia, South Africa and Australia,
(3) three groups for the Atlantic market with trade between Venezuela, 
the Gulf of Mexico, the American east coast and Northern Europe, and
(4) the American Pacific Coast.
It should be noted that the network includes the transport of crude oil as
well as commerce with already refined oil products so that oil producing
regions do not appear as separate communities.
%AK
This may be due to the limit in the detectability of smaller communities by
modularity optimization (Fortunato \& Barth\'elemy 2007), but does not affect the
relevance of the revealed ship traffic communities.
Because of the, by definition, higher transport intensity within 
communities, bioinvasive
spread is expected to be heavier between ports of the same community. 
However, in Fig.~\ref{Modularity} it becomes clear that there are no strict 
geographical barriers between communities.  Thus, spread between communities 
is very likely to occur even on small spatial scales by shipping or ocean 
currents between 
close-by
ports that belong to different communities.
  
\begin{figure}[tb] 
  \begin{center}
    \includegraphics[angle=0,width=1.0\textwidth,clip]{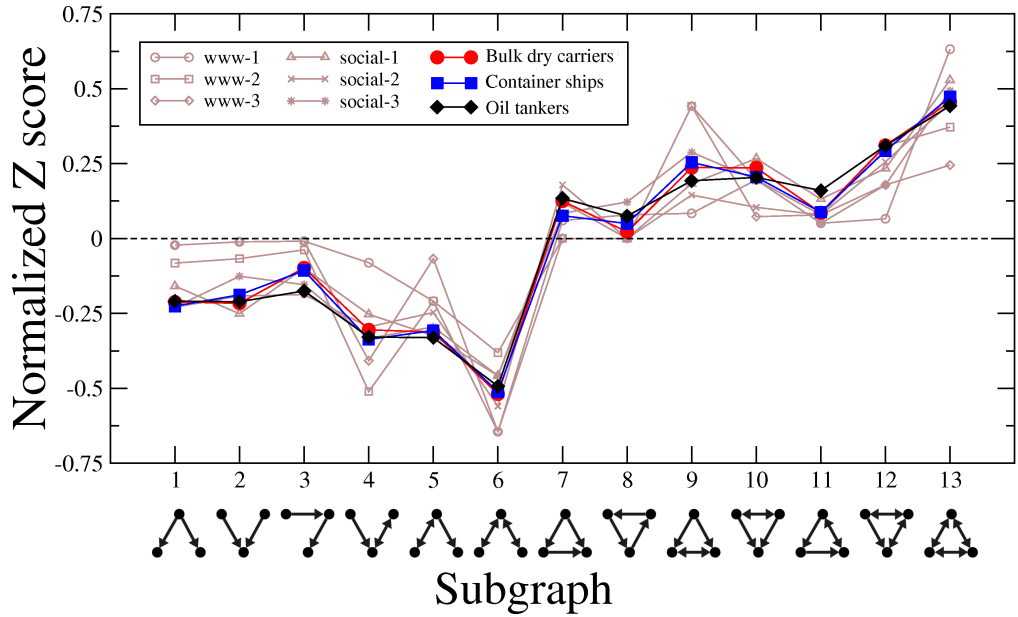}
    \caption{\label{motif} Motif distributions of the three main
      cargo fleets. 
      A positive (negative) normalized $Z$ score indicates that a motif is 
      more (less) frequent in the real network than in random networks with 
      the same degree sequence.  For comparison, we overlay the $Z$ scores 
      of the World Wide Web and social networks.  The agreement suggests 
      that the ship networks fall in the same superfamily of 
      networks (Milo \etal 2004).  The motif distributions of the fleets 
      are maintained even when 25\%, 50\% and 75\% of the weakest
      connections are removed.
    }
  \end{center}
\end{figure}

Despite the differences between the three main cargo fleets, there is one 
unifying feature: their motif distribution (Milo \etal 2002).  
Like most previous studies, we focus here on the occurrence of three-node 
motifs and present their normalized $Z$ score, a measure for their abundance 
in a network (Fig.~\ref{motif}).
Strikingly, the three fleets have practically the same motif distribution.  
In fact, the $Z$ scores closely resemble those found in the World Wide Web 
and different social networks which were conjectured to form a superfamily 
of networks (Milo \etal 2004).
This superfamily displays many transitive triplet interactions (i.e. if  
$X \to Y$ and $Y \to Z$, then $X \to Z$); for example, the overrepresented 
motif 13 in Fig.~\ref{motif}, has six such interactions.  
Intransitive motifs, like motif 6, are comparably infrequent.  
The abundance of transitive interactions in the ship networks indicates that 
cargo can be transported both directly between ports as well as via several
intermediate ports.
Thus, the high clustering and redundancy of links (robustness to link 
failures) appears not only in the GCSN but also in the three subnetworks. 
The similarity of the motif distributions to other humanly optimized networks 
underlines that  
cargo trade, like social networks and the World Wide Web, depends crucially on 
human interactions and information exchange. 
While advantageous for the robustness of trade, the clustering of links as 
triplets also has an unwanted side effect: in general, the more clustered a
network, the more vulnerable it becomes to the global spread of alien
species, even for low invasion probabilities (Newman 2003b).

\section{Network trajectories}
Going beyond the network perspective, the data base also provides 
information about the movement characteristics per individual ship
(Table~\ref{tab:trajs}).
The average number of distinct ports per ship $\langle N\rangle$ does not 
differ much between different ship classes, but container ships call much 
more frequently at ports than bulk dry carriers and oil tankers. 
This difference is explained by the characteristics and operational mode of 
these ships.  
Normally, container ships are fast (between 20 and 25 knots) and spend less 
time ($1.9$ days on average in our data) in the port for cargo operations.  
By contrast, bulk dry carriers and oil tankers move more slowly (between 13 
and 17 knots) and stay longer in the ports (on average $5.6$ days for bulk 
dry carriers, $4.6$ days for oil tankers).

\begin{figure}[t] 
  \begin{center}
    \includegraphics[angle=0,width=0.9\textwidth,clip]{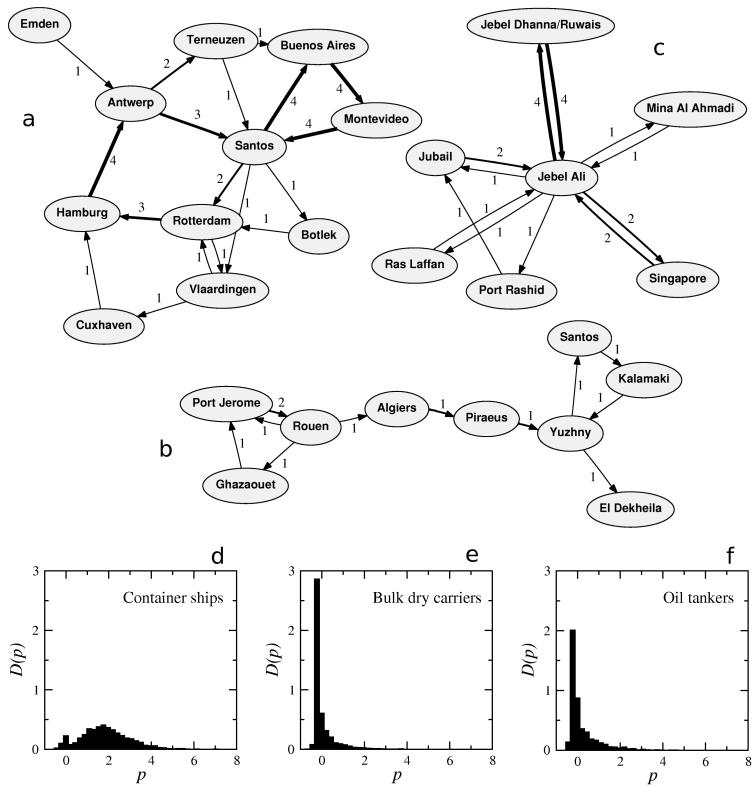}
    \caption{\label{distributions_p} Sample trajectories of (a) a container
      ship with a regularity index $p=2.09$, (b) a bulk dry carrier, 
      $p=0.098$, (c) an oil tanker, $p=1.027$.  In the three trajectories, 
      the numbers and the line thickness
      indicate the frequency of journeys 
      on each link.
      (d)-(f) Distribution of $p$ for the three main fleets.}
  \end{center}
\end{figure}

The speed at sea and of cargo handling, however, is not the only operational
difference.  The topology of the trajectories also differs substantially.  
Characteristic sample trajectories for each ship type are presented in 
Fig.~\ref{distributions_p}a-c.  
The container ship (Fig.~\ref{distributions_p}a) travels on some of the 
links several times during the study period whereas the bulk dry carrier 
(Fig.~\ref{distributions_p}b) passes almost every link exactly once.  The 
oil tanker (Fig.~\ref{distributions_p}c) commutes a few times between some 
ports, but by and large also serves most links only once.
    
We can express these trends in terms of a ``regularity index'' $p$ that
quantifies how much the frequency with which each link is used deviates from 
a random network.  
Consider the trajectory of a ship calling $S$ times at $N$ distinct ports and
travelling on $L$ distinct links.  
We compare the mean number of journeys per link $f_{real}=S/L$ to the average 
link usage $f_{ran}$ in an ensemble of randomized trajectories with the same 
number of nodes $N$ and port calls $S$. 
To quantify the difference between real and random trajectories we calculate 
the $Z$ score $p=(f_{real} - f_{ran})/\sigma$ (where $\sigma$ is the standard 
deviation of $f$ in the random ensemble).
If $p=0$, the real trajectory is indistinguishable from a random walk,
whereas larger values of $p$ indicate that the movement is more regular.
Figures \ref{distributions_p}d-f
present the distributions of the regularity
index $p$ for the different fleets.  
For container ships, $p$ is distributed broadly around $p\approx 2$,
thus supporting our earlier observation that most container ships 
provide regular services between ports along their way.  
Trajectories of bulk dry carriers and oil tankers, on the other hand, appear 
essentially random with the vast majority of ships near $p=0$.

\section{Approximating traffic flows using the gravity model}

In this article, we view global ship movements as a network based on
detailed arrival and departure records.
Until recently, surveys of seaborne trade had to rely on far less data:
only the total number of arrivals at some major ports were publicly 
accessible, but not the ships' actual paths (Zachcial \& Heideloff 2001).
Missing information about the frequency of journeys, thus, had to be
replaced by plausible assumptions, the gravity model being the most popular
choice.
It posits that trips are, in general, more likely between nearby ports
than between ports far apart.
If $d_{ij}$ is the distance between ports $i$ and $j$, the decline in
mutual interaction is expressed in terms of a distance deterrence function
$f(d_{ij})$.
The number of journeys from $i$ to $j$ then takes the form
$F_{ij} = a_i b_j O_i I_j f(d_{ij})$, where $O_i$ is the total number of
departures from port $i$ and $I_j$ the number of arrivals at
$j$ (Haynes \& Fotheringham 1984).
The coefficients $a_i$ and $b_j$ are needed to ensure $\sum_j F_{ij} =
O_i$ and $\sum_i F_{ij} = I_j$.

\begin{figure}[t]
  \begin{center}
    \includegraphics[width=1.0\textwidth,clip]{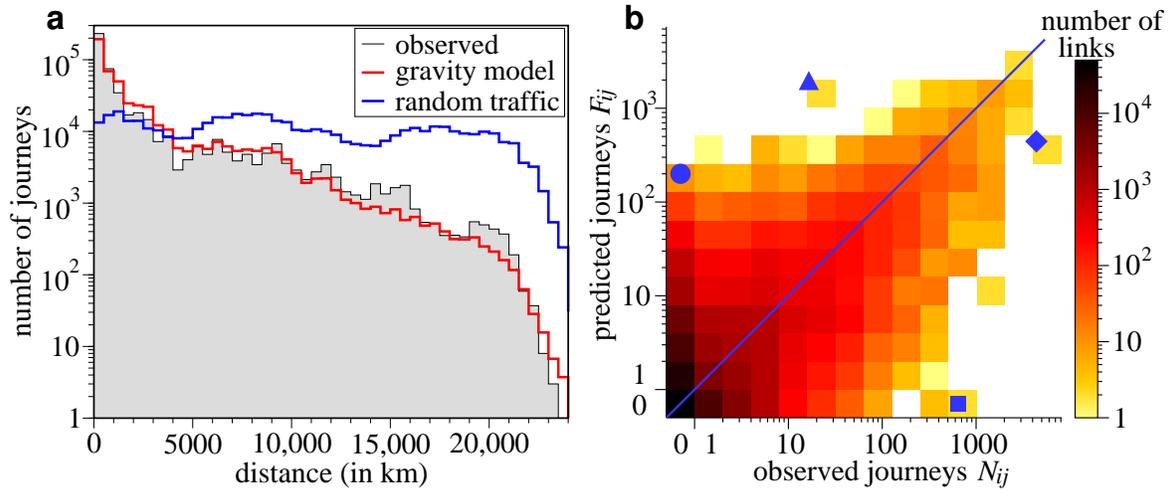}
    \caption{\label{gravity}
      (a) Histogram of port-to-port distances travelled in the GCSN
      (navigable distances around continents as indicated in
      Fig.~\ref{full_netw}).  We overlay the predictions of two different
      models. The gravity model (red), based on information about
      distances between ports and total port calls, gives a much better fit 
      than a simpler model (blue) which only fixes the total number of
      journeys. (b) Count of port pairs with $N_{ij}$ observed and $F_{ij}$
      predicted journeys.  The flows $F_{ij}$ were calculated with the 
      gravity model (rounded to the nearest integer).  
      Some of the worst outliers are highlighted in blue. $\circ$: Antwerp 
      to Calais ($N_{ij}=0$ vs.~$F_{ij}=200$). $\bigtriangleup$: Hook of 
      Holland to Europoort ($16$ vs.~$1895$). $\diamond$: Calais to Dover 
      ($4392$ vs.~$443$). $\Box$: Harwich to Hook of Holland ($644$ vs.~$0$).}
  \end{center}
\end{figure}

How well can the gravity model approximate real ship traffic?  We choose a 
truncated power law for the deterrence function, 
$f(d_{ij}) = {d_{ij}}^{-\beta}\exp(-d_{ij}/\kappa)$.
The strongest correlation between model and data is obtained for $\beta=0.59$
and $\kappa=4900$ km (see Electronic Supplementary Material).
At first sight, the agreement between data and model appears indeed
impressive.
The predicted distribution of travelled distances (Fig.~\ref{gravity}a)
fits the data far better than a simpler non-spatial model that preserves the
total number of journeys, but assumes completely random origins and
destinations.

A closer look at the gravity model, however, reveals its limitations.
In Fig.~\ref{gravity}b we count how often links with an observed number of
journeys $N_{ij}$ are predicted to be passed $F_{ij}$ times.
Ideally all data points would align along the diagonal $F_{ij}=N_{ij}$,
but we find that the data are substantially scattered.
Although the parameters $\beta$ and $\kappa$ were chosen to minimize the
scatter, the correlation between data and model is only moderate
(Kendall's $\tau=0.433$).
In some cases, the prediction is off by several thousand journeys per
year.

Recent studies have used the gravity model to pinpoint the ports and routes
central to the spread of invasive species (Drake \& Lodge 2004,
Tatem \etal 2006).  The model's shortcomings pose the question how reliable 
such predictions are.  
For this purpose, we investigated a dynamic model of ship-mediated bioinvasion 
where the weights of the links are either the observed traffic flows or the 
flows of the gravity model.  

We follow previous epidemiological studies
(Rvachev \& Longini 1985, Flahault \etal 1988, Hufnagel \etal 2004, Colizza 
\etal 2006) in viewing the spread on the network as a metapopulation process 
where the population dynamics on the nodes are coupled by transport on the 
links.
In our model, ships can transport a surviving population of an invasive species 
with only a small probability $p_{trans}=1\%$ on each journey between two successively
visited ports.
The transported population is only a tiny fraction $s$ of the
population at the port of origin.
Immediately after arriving at a new port, the species experiences strong 
demographic fluctuations which lead in most cases to the death of the imported
population.
If however the new immigrants beat the odds of this ``ecological roulette'' 
(Carlton \& Geller 1993) and establish, 
the population $P$ grows 
rapidly following the
stochastic
logistic equation 
$\frac{dP}{dt} = rP(1-P) + \sqrt{P}\xi(t)$
with growth rate $r=1/\text{year}$ and Gaussian white noise $\xi$.
For details of the model, we refer to the Electronic Supplementary Material.

\begin{figure}[t]
  \begin{center}
    \includegraphics[width=0.7\textwidth,clip]{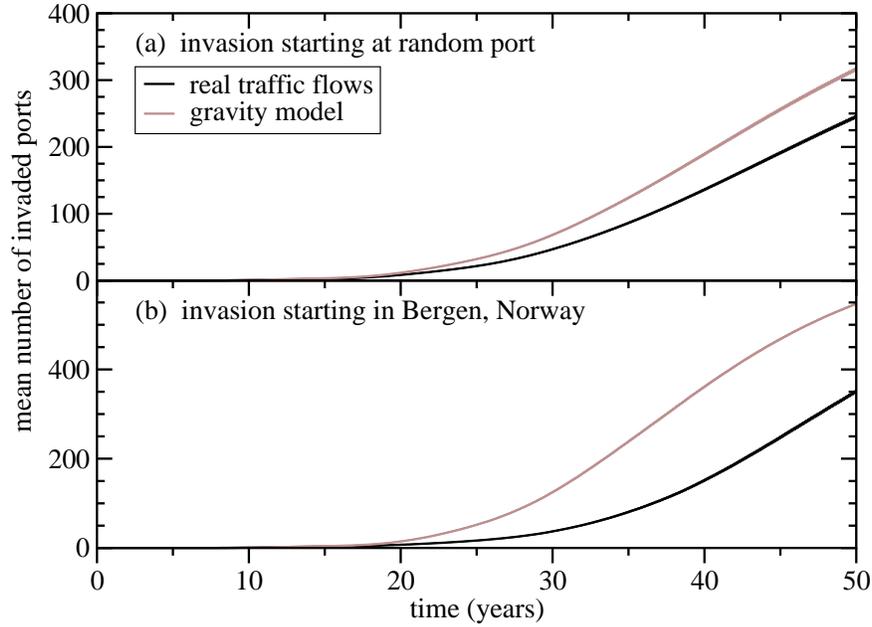}
    \caption{\label{metapop}
      Results from a stochastic population model for the spread of an invasive 
      species between ports.  
      (a) The invasion starts from one single, randomly chosen port.  
      (b) The initial port is fixed as Bergen (Norway), an example of a
      well-connected port (degree $k=49$) which is not among the central
      hubs.
      The rate of journeys from port $i$ to $j$ per year is assumed to be 
      $N_{ij}$ (real flows from the GCSN) or $F_{ij}$ (gravity model).
      Each journey has a small probability of transporting a tiny fraction 
      of the population from origin to destination.
      Parameters were adjusted ($r=1/\text{year}$, $p_{trans}=0.01$,  $s=4\cdot10^{-5}$) to yield a per-ship-call probability of 
      initiating invasion of $\approx 4.4\cdot10^{-4}$ (Drake \& Lodge 2004, 
      see Electronic Supplementary Material for details).  
      Plotted are the cumulative numbers of invaded ports (population 
      number
      larger
      than half the carrying capacity) averaged over (a) $14,000$, (b)
      $1000$ simulation runs (standard error equal to line thickness).
    }
  \end{center}
\end{figure}

Starting from a single port at carrying capacity $P=1$, we model contacts 
between ports as Poisson processes with rates $N_{ij}$ (empirical data) 
or $F_{ij}$ (gravity model).  
As shown in Fig.~\ref{metapop}a, the gravity model systematically 
overestimates the spreading rate, and the difference can become particularly 
pronounced for ports which are well-connected, but not among the central hubs 
in the network (Fig.~\ref{metapop}b).  
Comparing typical sequences of infected ports, we find that the invasions 
driven by the real traffic flows tend to be initially confined to smaller 
regional ports, whereas in the gravity model the invasions quickly reach the 
hubs. 
%MG 16 Dec 09: Following the referee's comments, more about the correlation
%  in real and gravity network.
%The gravity model apparently erases too many details of a hierarchical 
%structure present in the real network.
The total out- and in-flows at the ship journeys' origin and departure ports,
respectively, are indeed more strongly positively correlated in reality than 
in the model ($\tau = 0.157$ vs.~ $0.047$).
The gravity model thus erases too many details of a hierarchical structure 
present in the real network.
That the gravity model eliminates most correlations, is also plausible from
simple analytic arguments, see Electronic Supplementary Material for details.
The absence of strong correlations makes the gravity model a suitable null 
hypothesis if the correlations in the real network are unknown, but several 
recent studies have shown that correlations play an important role in
spreading processes on networks (e.g. Newman 2002, Bogu\~na \& Pastor-Satorras
2002).
Hence, if the correlations are known, they should not be ignored.

%While the qualitative difference between AIS data and the gravity model proved
%stable against changes in the parameter settings and even different population
%models, the time scale of the invasion is much less predictable.  
While we observed that the spreading rates for the AIS data were consistently 
slower than for the gravity model even when different parameters or population 
models were considered, the time scale of the invasion is much less 
predictable. 
%\MG
The assumption that only a small fraction of invaders succeed outside their
native habitat appears realistic (Mack \etal 2000).  
Furthermore, the parameters in our model were adjusted so that the 
per-ship-call probability of initiating invasion is approximately 
$4.4\cdot 10^{-4}$, a rule-of-thumb value stated by Drake \& Lodge (2004).  
Still, too little is empirically known to pin down individual parameters with 
sufficient accuracy to give more than a qualitative impression.
It is especially difficult to predict how a potential invader reacts to the 
environmental conditions at a specific location.
Growth rates certainly differ greatly between ports depending on factors
such as temperature or salinity, 
with respect to the habitat requirements of the invading organisms.
Our results should, therefore, be regarded as 
one of many different conceivable scenarios.
A more detailed study of bioinvasion risks 
based on the GCSN 
is currently underway (Seebens
\& Blasius 2009).

\section{Conclusion}
We have presented a study of ship movements based on AIS records.  
Viewing the ports as nodes in a network linked by ship journeys, we found 
that global cargo shipping, like many other complex networks investigated
in recent years, possesses the small world property as well as broad 
degree and weight distributions.
Other features, like the importance of canals and the grouping of ports into
regional clusters, are more specific to the shipping industry.
An important characteristic of the network are differences in the movement
patterns of different ship types. 
Bulk dry carriers and oil tankers tend to move in a less regular manner 
between ports than container ships.
This is an important result regarding the spread of invasive species 
because bulk dry carriers and oil tankers often sail empty and therefore 
exchange large quantities of ballast water.
The gravity model, which is the traditional approach to forecasting 
marine biological invasions, captures some broad trends of global cargo trade, 
but for many applications its results are too crude.
Future strategies to curb invasions will have to take more details into 
account.
The network structure presented in this article can be viewed as a first
step in this direction.

\begin{table*}[t]
  \begin{tabular*}{\hsize}{@{\extracolsep{\fill}}lccccccccccccc}
    Ship class &  ships &  MGT & $n$ & $\langle k \rangle$ & $C$  & $\langle l \rangle$ & $\langle J \rangle $ & $\mu$ & $\eta$ &
    $\langle N \rangle$ & $\langle L \rangle$ & $\langle S \rangle$ & $\langle p \rangle$ \cr
    \hline 
    Whole fleet      & 16363 & 664.7 & 951 & {\bbb 76.4} & 0.49 & 2.5  & 13.57 & 1.71 & 1.02 & 10.4 & 15.6 & 31.8 & 0.63\\[2pt]
    Container ships   & 3100 & 116.8 & {\bbb 378} & 32.4 & 0.52 & 2.76 & {\bbb 24.25} & {\bbb 1.42} & 1.05 &  11.2 & {\bbb 21.2 } & {\bbb 48.9} & {\bbb 1.84}\\
    Bulk dry carriers	& 5498 & 196.8 & 616 & 44.6 & 0.43 & 2.57 &  4.65 & 1.93 & 1.13 &  8.9  & 10.4 & 12.2 &  0.03\\
    Oil tankers 	& 2628 & 178.4 & 505 & 33.3 & 0.44 & 2.74 &  5.07 & 1.73 & 1.01 &  9.2  & 12.9 & 17.7 & 0.19\\
    \hline  \\[-2mm]
  \end{tabular*}
  \caption{\label{tab:trajs} 
    Characterization of different subnetworks.
    Number of ships, total gross tonnage $[10^6$ GT$]$ and number of 
    ports $n$ in each subnetwork; together with network characteristics: 
    mean degree $\langle k \rangle$, clustering coefficient $C$, mean 
    shortest path length $\langle l \rangle$, mean journeys per link 
    $\langle J \rangle $, power-law exponents $\mu$ and $\eta$; 
    and trajectory properties: average number of distinct ports 
    $\langle N \rangle$, links $\langle L \rangle$, port calls 
    $\langle S \rangle$ per ship and regularity index $\langle p \rangle$.
    Some notable values are highlighted in bold.
  }
\end{table*}

\begin{acknowledgments}
We thank B. Volk, K.~H. Holocher, A. Wilkinson, J.~M. Drake and H. Rosenthal 
for stimulating discussions and helpful suggestions.
We also thank Lloyd's Register Fairplay  for providing their shipping data base.
This work was supported by German VW-Stiftung
and BMBF.
\end{acknowledgments}

{\bf Supplementary information} is linked to the online version of the paper at Journal of the Royal Society Interface

\end{document}